\documentclass[12pt]{article}

\usepackage[percent]{overpic}
\usepackage{float}
\usepackage{pgfplots}
\usepackage{placeins}
\usepackage{amsmath}
\usepackage{amsthm}
\usepackage{color}
\usepackage{amssymb}
\usepackage{mathtools}
\usepackage{subfigure}
\usepackage{multirow}
\usepackage{epsfig}
\usepackage{listings}
\usepackage{enumitem}
\usepackage{rotating,tabularx}
\usepackage{graphicx}
\usepackage{graphics}
\usepackage{epstopdf}
\usepackage{longtable}
\usepackage[pdftex]{hyperref}
\usepackage{epigraph}
\usepackage{xspace}
\usepackage{amsfonts}
\usepackage{eurosym}
\usepackage{ulem}
\usepackage{footmisc}
\usepackage{comment}
\usepackage{setspace}
\usepackage{geometry}
\usepackage{caption}
\usepackage{pdflscape}
\usepackage{array}
\usepackage[round]{natbib}
\usepackage{booktabs}
\usepackage{dcolumn}
\usepackage{mathrsfs}

\usepackage{enumitem}
\usepackage{tikz}
\usetikzlibrary{decorations.pathreplacing}

\usepackage{xcolor}
\hypersetup{
colorlinks,
linkcolor={blue!50!black},
citecolor={blue!50!black},
urlcolor={blue!50!black}}

\newcommand{\bc}{\begin{center}}
\newcommand{\ec}{\end{center}}

\newtheorem{definition}{Definition}

\newcolumntype{d}[1]{D{.}{.}{#1}} 

\newcolumntype{L}[1]{>{\raggedright\let\newline\\arraybackslash\hspace{0pt}}m{#1}}
\newcolumntype{C}[1]{>{\centering\let\newline\\arraybackslash\hspace{0pt}}m{#1}}
\newcolumntype{R}[1]{>{\raggedleft\let\newline\\arraybackslash\hspace{0pt}}m{#1}}

\newcommand{\elabel}[1]{\label{eq:#1}}
\newcommand{\eref}[1]{Eq.~(\ref{eq:#1})}

\newcommand{\sref}[1]{Sec.~\ref{sec:#1}}

\newcommand{\Dref}[1]{Definition~\ref{def:#1}}

\newcommand{\Cref}[1]{Corollary~\ref{coro:#1}}
\newcommand{\cref}[1]{Cor.~\ref{coro:#1}}

\newcommand{\ie}{{\it i.e.}\xspace}
\newcommand{\eg}{{\it e.g.}\xspace}

\newcommand{\cf}{{\it c.f.}\xspace}

\newcommand{\flabel}[1]{\label{fig:#1}}
\newcommand{\fref}[1]{Fig.~\ref{fig:#1}}
\newcommand{\Fref}[1]{Figure~\ref{fig:#1}}

\newcommand{\be}{\begin{equation}}
\newcommand{\ee}{\end{equation}}
\newcommand{\bea}{\begin{eqnarray}}
\newcommand{\eea}{\end{eqnarray}}

\newcommand{\bi}{\begin{itemize}}
\newcommand{\ei}{\end{itemize}}

\newcommand{\Dt}{\Delta t}
\newcommand{\Dx}{\Delta x}

\newcommand{\del}{D}
\newcommand{\hor}{H}
\newcommand{\subhead}[1]{\mbox{}\newline\textbf{#1}\newline}

\numberwithin{equation}{section}

\begin{document}

\begin{titlepage}
\title{Microfoundations of Discounting}
\author{Alexander T. I. Adamou\footnote{London Mathematical Laboratory,~\url{a.adamou@lml.org.uk}} \and Yonatan Berman\footnote{London Mathematical Laboratory,~\url{y.berman@lml.org.uk}} \and  Diomides P. Mavroyiannis\footnote{Universit\'{e} Paris-Dauphine,~\url{diomides.mavroyiannis@dauphine.eu}} \and Ole B. Peters\footnote{London Mathematical Laboratory and Santa Fe Institute,~\url{o.peters@lml.org.uk}}\,\, \thanks{We thank Ryan Singer for insightful discussions that led to an early outline for this manuscript. We are also grateful to Matthew Gentry, and to seminar participants at the 15th European Meeting on Game Theory, 2019 International Conference on Public Economic Theory, the Danish Research Centre for Magnetic Resonance, Royal Holloway, University of London, the University of Cyprus, and Universit\'{e} Paris-Dauphine. We thank Baillie Gifford for sponsoring the Ergodicity Economics program at the London Mathematical Laboratory.}}
\date{\today}
\maketitle


\begin{abstract}
\noindent An important question in economics is how people choose between different payments in the future. The classical normative model predicts that a decision maker discounts a later payment relative to an earlier one by an exponential function of the time between them. Descriptive models use non-exponential functions to fit observed behavioral phenomena, such as preference reversal. Here we propose a model of discounting, consistent with standard axioms of choice, in which decision makers maximize the growth rate of their wealth. Four specifications of the model produce four forms of discounting -- no discounting, exponential, hyperbolic, and a hybrid of exponential and hyperbolic -- two of which predict preference reversal. Our model requires no assumption of behavioral bias or payment risk.

\end{abstract}
\setcounter{page}{0}
\thispagestyle{empty}
\end{titlepage}
\pagebreak \newpage

\section{Introduction}\label{sec:introduction}

\subsection{Background}\label{sec:background}

In economics and psychology, temporal discounting -- or, simply, discounting -- is a paradigm of how decision makers choose between rewards available at different times in the future. We write here of people and money payments, noting that discounting is also studied in other animals and for other reward types. The basic observation to be explained is this: for two payments of equal size, people prefer typically the earlier payment to the later one. In the discounting paradigm, the later payment is discounted relative to the earlier payment by multiplying it by a function of the time period between the payments, called the \textit{delay}. This operation expresses the later payment as an equivalent payment at the earlier time, to be compared with the earlier payment actually on offer.

Why do people assign lower values to payments further in the future? One plausible answer is that a later payment is less likely to be received, because there is more time for something to go wrong with it. In other words, delay increases risk. Another is that, for equal payments, the later one corresponds to a lower growth rate which, if sustained over time, would result in being poorer. Modern treatments of discounting in economics tend to follow risk-based reasoning, while there is a more even split between risk and rate interpretations in psychology.

This paper studies the microfoundations of discounting using the rate interpretation in a riskless setting. In our model, a decision maker chooses between two known and different payments to be received at known and different times, such that the growth rate of her wealth is maximized. Our model assumes no behavioral bias and does not violate standard axioms of choice. It predicts a range of situation-dependent discount functions, including those documented in the discounting literature.

This literature abounds with models \citep{CohenETAL2019}. In some, theoretical considerations are used to construct the decision maker's maximand, from which the discount function is derived. Such models are ``normative'' in that they say what a decision maker should do if she wants to act optimally in the sense specified. In other models, the discount function is chosen to fit empirical data, with theoretical justification sought {\it post hoc} or not at all. These models are ``descriptive'' in that they predict what decision makers actually do, regardless of whether it is in any sense optimal.

The main normative model in economics is exponential discounting, in which the discount function decays exponentially with the delay \citep{Samuelson1937}. For money payments, this is derived straightforwardly: either by a no-arbitrage argument, assuming payments are guaranteed and the earlier payment can be invested during the delay to earn compound interest at a riskless rate; or by assuming the later payment has a constant hazard rate 
during the delay and the decision maker maximizes the expected payment \citep{Kacelnik1997}.

Exponential discounting is not descriptive. Experiments on human and non-human animals suggest that payments can be discounted more for shorter delays and less for longer delays than is well described by fitting the rate parameter of an exponential function. Furthermore, subjects exhibit a behavior known as preference reversal, where they switch from preferring the later to the earlier payment as time passes. Specifically, the switch happens as the time to the earlier payment -- which we call the \textit{horizon} -- gets shorter, while the delay between the payments stays the same \citep[p.~288]{KerenRoelofsma1995}. Preference reversal is never predicted by standard exponential discounting \citep[Fig~2]{GreenMyerson1996}. The primary evidence against the main normative model is summarized by \citet{MyersonGreen1995} and references therein. In response to its falsification, descriptive models are proposed with discount functions better able to fit observations.

A widely-used descriptive model is hyperbolic discounting, where the discount function is a hyperbola in the delay. The function has one free parameter, known as the degree of discounting, which determines its steepness. Fitting this parameter to experimental data is more efficient than fitting an exponential function, both at group level and for individuals \citep{MyersonGreen1995}. Furthermore, preference reversal is compatible with this model \citep[Fig.~2]{GreenMyerson1996}.

\citet{Kacelnik1997} remarks on this divergence between normative and descriptive models, noting that the hyperbola ``is not strongly explanatory because it did not emerge from an {\it a priori} analysis but purely from its power to describe data efficiently. In contrast, because of the strong appeal of the {\it a priori} argument favoring exponential discounting, several re-elaborations have been made to rescue the rationale that led to it.''\footnote{Said succinctly, a normative-only model provides rationale without fit, and a descriptive-only model provides fit without rationale. Ideally, models of discounting, as of other behavioral phenomena, would provide both and the distinction would be redundant.} Most such attempts to adapt the normative model introduce payment uncertainty, which we discuss in \sref{literature}. Another approach, favored in behavioral economics, is to present non-exponential discounting as a cognitive bias -- a deviation from optimal behavior -- to be documented and quantified in mathematical functions that encode human psychology \citep{LoewensteinPrelec1992,Laibson1997}.

\subsection{Our model -- growth rate maximization\label{sec:ourmodel}}

Here we propose a model of temporal discounting compatible with hyperbolic discounting, in which neither payment risk nor behavioral bias are assumed. We study the basic temporal choice problem in which a decision maker must choose between two known, certain, and different payments to be made at known, certain, and different future times. In our model, she does so by comparing the growth rates of wealth associated with each option.

The temporal choice problem is underspecified. We specify it fully by introducing: the wealth dynamics, treating specifically additive and multiplicative cases; and the time frame of the decision, meaning the period over which it is appropriate to compute growth rates. 

Depending on the specification, our model predicts four different forms of discounting: no discounting; exponential; hyperbolic; and a hybrid of exponential and hyperbolic. This is not an exhaustive list -- other dynamics would produce other forms of discounting. Two of the discount functions nested in our model, hyperbolic and hybrid, are compatible with preference reversal.

The hybrid discount function depends not only on the delay and the horizon, but also on the decision maker's wealth and background growth rate. This produces a richer set of predicted behaviors than other specifications. One prediction is that decision makers can switch from preferring the earlier to the later payment as their wealth increases. In other words, richer people discount less steeply, consistent with empirical findings \citep{GreenETAL1996,EpperETAL2018}. Another prediction is that, for small payments, discounting is close to hyperbolic for short delays and close to exponential for long delays.

The main contribution of this paper is the prediction of non-exponential discounting and preference reversal in a normative model that does not violate standard axioms of choice \citep{vonNeumannMorgenstern1944}. Our model assumes neither bias nor dynamic inconsistency \citep[p.~3]{CohenETAL2019} in the decision maker's behavior. At all times she prefers the option with the highest growth rate. In some specifications this translates into a reversal of preference between {\it payments} (not growth rates). In our perspective this reflects a change of circumstances and not of mind. The fundamental preference -- for faster growth -- never reverses.

Moreover, this paper marks a shift from psychological to circumstantial explanations of discounting. We predict that changes in the discount function arise from changes in wealth dynamics and time frame, which are properties not of the decision maker but of her circumstances. When these circumstances are included in the formalism, a single behavioral model -- a single maximand -- is capable of predicting a range of observed behaviors. Since psychological risk preferences, encoded in idiosyncratic utility functions, do not appear in our model, we sidestep recent controversy in the literature about the suitability of experiments involving money payments to test models of utility-of-consumption flows \citep{CohenETAL2019}. Such experiments are able to falsify our model and are planned.

The paper is organized as follows. \sref{model} sets out the temporal choice problem and our decision model. In \sref{results} we present different specifications of the problem and describe how a decision maker discounts future payments in each of them. We conclude in \sref{discussion}.

\subsection{Related literature\label{sec:literature}}

This paper follows the tradition of adapting the normative discounting model to make it consistent with observations, \ie to make it descriptive \citep{Kacelnik1997}. Our strategy is to postulate the growth rate of wealth as the decision maker's maximand. Computing this requires information about wealth dynamics and time frame, about which the basic temporal choice problem is silent.

Another strategy is to leave the maximand -- usually an expected payment or utility gain -- unchanged and introduce uncertainty in the amount or timing of the payment. The uncertainty is chosen so that the effective discount function takes a form known to fit the data. Adding risk can be viewed as another way of resolving the underspecification of the temporal choice problem. This approach is questionable because, whereas dynamics are unspecified in the problem, uncertainty is explicitly absent -- a choice between certain payments at certain times. Payment risk is a sound microfoundation when the uncertainty absent from the problem formulation is important in reality. Such situations are plausible and likely widespread. For example, a predator declining a small but readily-caught prey to search for something more filling risks catching nothing at all. However, adding payment uncertainty to generate, say, hyperbolic discounting is not a general prescription. It fails when the real payment risk is negligible, as envisaged in the problem and presented in experiments, \eg \citet{MyersonGreen1995}.

Furthermore, to recover the hyperbola as the discount function, specific forms of uncertainty or other adaptations are required. \citet{GreenMyerson1996} point out that an expected payment model (`risk neutrality') with a hyperbolic hazard rate predicts hyperbolic discounting. They note also that the exponential function can be made consistent with observed behavior, including preference reversal, by allowing its rate parameter to vary with payment size. \citet{Sozou1998} treats an expected payment model with an uncertain hazard rate, about which the decision maker learns through Bayesian updating. If the prior distribution of the hazard rate is exponential, then hyperbolic discounting is again obtained. \citet{dasgupta2005uncertainty} also assume risk neutrality but keep the hazard rate constant. To recover hyperbolic discounting they introduce the possibility of payment occurring before the anticipated time.

Such adaptations lead to a loss of generality. They make statements of the type: `if there is uncertainty in a future payment, and if it takes this specific form, then the discount function is a hyperbola.' Such {\it ad hoc} assumptions reduce the generality of risk-based models further, since they are useful only when the real risk is of a particular nature.

The strategy we follow leaves payment certainty alone and changes the maximand. It is long established in biology and psychology that hyperbolic discounting is consistent with maximizing the rate of change of resources in a model of additive payments. This insight, traced back to the predation model of \citet{Holling1959}, is recognized in studies of human discounting \citep{MyersonGreen1995,Kacelnik1997,Sozou1998}. \citet[Fig.~2]{Kacelnik1997} offers a pictorial representation, similar to ours in \sref{results}, of how rate maximization predicts preference reversal. In the rate interpretation, the degree of discounting is no longer a free psychological parameter. Rather it is constrained to be the reciprocal of the horizon \citep{MyersonGreen1995}. This is a testable prediction.

This paper extends this strand of the literature by setting the decision maker's maximand as the growth rate of resources under general dynamics. We generalize further by allowing the time frame of the decision -- over which growth rates are computed -- to be the period from the decision to either the chosen or the later payment. This captures circumstances akin to opportunity costs, specifically whether receiving the earlier payment frees the decision maker to pursue other payments.

Our work contributes to the growing field of ergodicity economics \citep{Peters2019b} in which decision makers maximize the long-time growth rate of resources, rather than expectation values of psychologically-transformed resource flows (under, in prospect theory, psychologically-transformed probability measures). This study joins recent evidence of strong dependence on wealth dynamics of human decisions under uncertainty \citep{MederETAL2019} and may be used to design similar experiments without uncertainty.

\section{Theoretical Framework}\label{sec:model}

\subsection{Problem definition}\label{sec:problem}

We begin by formalizing a basic riskless temporal choice problem, where discounting is used to express a later payment as an equivalent payment at an earlier time. We define a {\it Riskless Intertemporal Payment Problem} (RIPP):

\begin{definition}[Riskless Intertemporal Payment Problem]


A Riskless Intertemporal Payment Problem (RIPP) is a comparison between two vectors, $a \equiv \left(t_a,\Dx_a\right) , b \equiv \left(t_b,\Dx_b\right)$. A decision maker at time $t_0$ with wealth $x\left(t_0\right)$ must choose between two future cash payments, whose amounts and payment times are known with certainty. The two options are:
\begin{enumerate}
\item[$a$.] an earlier payment of $\Dx_a$ at time $t_a>t_0$; and
\item[$b$.] a later payment of $\Dx_b$ at time $t_b>t_a$.
\end{enumerate}
\end{definition}

A criterion for choosing $a$ or $b$ is required. Here we explore what happens if that criterion is maximization of the growth rate of wealth, \ie if $a$ is chosen when it corresponds to a higher growth rate of the decision maker's wealth than $b$, and \textit{vice versa}.

\subhead{Growth rates}
A growth rate is defined as the scale parameter of time in the growth function of wealth subject to dynamics. So, if wealth grows as $x\left(t-t_0\right)=f\left(g\left(t-t_0\right)\right)$, where $f$ denotes the growth function, then the growth rate is $g$. Different dynamics correspond to different growth functions and, therefore, to different forms of growth rate. We treat explicitly multiplicative and additive dynamics \citep{PetersGell-Mann2016}, noting that more general dynamics can be treated similarly \citep{PetersAdamou2018a}.

\subhead{Multiplicative dynamics}
Ignoring, for the moment, possible payments $\Dx_a$ and $\Dx_b$, a common assumption is that wealth grows exponentially in time. We label this dynamic as multiplicative. It corresponds to investing wealth in income-generating assets, where the income is proportional to the amount invested. Wealth grows as
\be
x\left(t\right) = x\left(t_0\right) e^{g \left(t - t_0\right)}\,,
\ee
and the scale parameter of time in the exponential function is $g$. This growth rate, $g$, resembles an interest rate or a rate of return on investment. Its dimension is the reciprocal of time, \eg 5\% per year.

\subhead{Additive dynamics}
Another possibility is additive dynamics, where wealth grows linearly in time. This corresponds to situations where investment income is negligible and wealth changes by net flows that do not depend on wealth itself. In this case wealth grows as
\be
x\left(t\right) = x\left(t_0\right) + g \left(t - t_0\right)\,,
\ee
and the scale parameter of time in the linear function is $g$. This growth rate, $g$, resembles a net savings rate from additive income, such as from labor earnings and welfare payments less consumption. It is measured in units of currency per unit of time, \eg \$5000 per year.

The functional form of the growth rate differs between the dynamics. The growth rate between time $t$ and $t+\Dt$ can be extracted from the expression for the evolution of wealth over that period. Under multiplicative dynamics it is
\be
g = \frac{\ln x\left(t+\Dt\right)-\ln x\left(t\right)}{\Dt}\,,
\ee
and under additive dynamics it is
\be
g = \frac{x\left(t+\Dt\right) - x\left(t\right)}{\Dt}\,.
\ee

The matching of growth rate with dynamics is crucial. An additive growth rate computed for wealth following a multiplicative process would vary with time, as would a multiplicative growth rate computed for additively-growing wealth. The correct growth rate extracts a stable parameter from the dynamics, allowing processes with the same type of dynamics to be compared.

Given the wealth dynamics, a RIPP implies two growth rates: $g_a$, associated with option $a$; and $g_b$, associated with option $b$. This permits defining growth-optimal behavior:

\begin{definition}[Growth-Optimal Preferences]

The preference relation $\succsim$ is growth-optimal if given the wealth dynamics, a decision time $t_0$, an initial wealth $x\left(t_0\right)$, and payments $a\equiv\left(t_a,\Dx_a\right)$ and $b\equiv\left(t_b,\Dx_b\right)$:
\begin{enumerate}
\item $a \succ b$ [`$a$ is preferred to $b$'] if and only if $g_a > g_b$ 
\item $a\sim b$ [`indifference between $a$ and $b$'] if and only if $g_a = g_b$
\item $a \prec b$ [`$b$ is preferred to $a$'] if and only if $g_a < g_b$
\end{enumerate}
\label{def:def1}
\end{definition}

In words, \Dref{def1} states that a growth-optimal decision maker prefers option $a$ if her wealth grows faster under this choice than under option $b$, and \textit{vice versa}. She is indifferent if the growth rates are equal. \Dref{def1} is consistent with the von Neumann-Morgenstern axioms \citep{vonNeumannMorgenstern1944}: completeness is satisfied by design. It also satisfies transitivity (see proof in Appendix~\ref{app:appA}). Independence and continuity are irrelevant since in this setup all the payments and times are certain.

%

\subsection{Model setup}

\Fref{basicsetup} illustrates a RIPP, corresponding to the basic question that arises in temporal discounting, \eg `would you prefer to receive \$100 tomorrow or \$200 in a month's time?' Growth rates depend on time increments, not times themselves, so it is useful to define the two fundamental time increments in the problem: the period from the decision to the earlier payment, called the \textit{horizon},
\be
\hor \equiv t_a-t_0\,;
\ee
and the period between the payments, called the \textit{delay},
\be
\del \equiv t_b-t_a\,.
\ee

The discount function is a function of the delay. It is the multiplicative factor which, when applied to the later payment, renders the decision maker indifferent, \ie
\be
\delta\left(\del\right) \equiv \left.\frac{\Dx_a}{\Dx_b}\right|_{a \sim b}\,.
\ee
Depending on the model specification, the discount function can also vary with other variables in the problem, such as the horizon, payment sizes, initial wealth, and any background growth rate.

\begin{figure}[!htb]
\centering
\begin{tikzpicture}
\draw[->,ultra thick] (-5,0)--(5,0) node[below,yshift=-4pt]{$Time$};
\draw[->,ultra thick] (-5,0)--(-5,6) node[left,xshift=-4pt]{$Wealth$};
\draw[-,ultra thick] (-4,-0.15)--(-4,0) node[below]{$t_0$};
\draw[-,ultra thick] (1,-0.15)--(1,0) node[below]{$t_a$};
\draw[-,ultra thick] (3,-0.15)--(3,0) node[below]{$t_b$};
\draw[-,ultra thick] (-5.15,0.75)--(-5,0.75) node[left]{$x\left(t_0\right)$};
\draw[-,ultra thick] (-5.15,2.75)--(-5,2.75) node[left]{$x\left(t_0\right) + \Dx_a$};
\draw[-,ultra thick] (-5.15,4.75)--(-5,4.75) node[left]{$x\left(t_0\right) + \Dx_b$};
\draw[-, dashed] (-5,0.75)--(5,0.75) ;
\draw[-, dashed] (-5,2.75)--(5,2.75) ;
\draw[-, dashed] (-5,4.75)--(5,4.75) ;
\draw[-, ultra thick] (1,0.75)--(1,2.75) ;
\draw[-, ultra thick] (3,0.75)--(3,4.75) ;
\draw [decorate,decoration={brace,amplitude=10pt,mirror},xshift=0pt,yshift=-5pt,thick](-3.8,0) -- (0.8,0) node [black,midway,yshift=-20pt]{$\text{Horizon, }\hor$};
\draw [decorate,decoration={brace,amplitude=10pt,mirror},xshift=0pt,yshift=-5pt,thick](1.2,0) -- (2.8,0) node [black,midway,yshift=-20pt]{$\text{Delay, }\del$};
\end{tikzpicture}
\caption{The basic setup of the model. A decision maker faces a choice at time $t_0$ between option $a$, which guarantees a payment of $\Dx_a$ at time $t_a$, and option $b$, which guarantees a payment of $\Dx_b>\Dx_a$ at time $t_b>t_a$. We define the time between the decision and the earlier payment as the {\it horizon}, $\hor\equiv t_a-t_0$; and the time between the two payments as the {\it delay}, $\del\equiv t_b-t_a$.}
\flabel{basicsetup}
\end{figure}
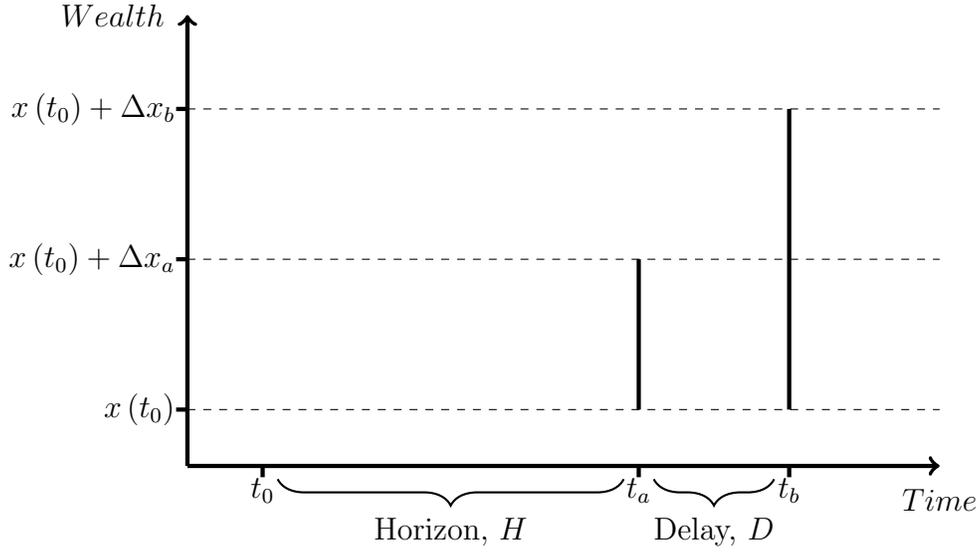

Despite its apparent simplicity, solving the temporal choice problem requires additional assumptions. In our model, two assumptions are needed. The first concerns the dynamics under which the decision maker's wealth grows, as discussed in \sref{problem}. This determines the appropriate form of the growth rate. The second assumption concerns what we call the \textit{time frame} of the decision, specifically whether a decision maker receiving the earlier payment at $t_a$ is free immediately to make her next decision, or whether she must wait until the later time $t_b$. This determines the appropriate time period for computing the growth rate under each option. Full specification allows the decision maker's maximand -- the growth rate of her wealth -- to be evaluated and her options compared. For concreteness we confine our attention to $\Dx_b > \Dx_a$, which is the most commonly considered dilemma.

We describe four different specifications of this basic setup. In each we calculate the growth rates of wealth, $g_a$ and $g_b$, associated with options $a$ and $b$. From this analysis we infer the discount function as
\be
\delta\left(\del\right) \equiv \left.\frac{\Dx_a}{\Dx_b}\right|_{g_a=g_b}\,,
\ee
\ie the ratio of earlier to later payment under the constraint that the growth rates corresponding to each are equal. The four specifications predict four forms of discounting -- no discounting, exponential, hyperbolic, and a hybrid of exponential and hyperbolic -- and, in some cases, preference reversal.

\section{Results}\label{sec:results}

\subsection{Specification}

We begin by describing the four different model specifications. Each specifies two aspects necessary to quantify the growth rate of wealth: the time frame of the decision; and the dynamics under which wealth evolves.

\subhead{Time frame}
A key aspect, often left unspecified or implicit in the literature, is whether receiving the earlier payment frees the decision maker to pursue the next payment. We treat this by specifying the time frame of the decision, which we illustrate with the following scenarios:
\begin{enumerate}
\item Denise is a day laborer. Every evening she looks at a job market forum and chooses a job for the next day. Jobs pay different wages and take different amounts of time, although always less than a day. Denise is paid as soon as she completes the job and goes home. She cannot do more than one job each day.
\item Fiona is a freelancer. She works on projects ranging from a few days to many months and she can only work on one project at a time. As soon as she finishes a project, she gets paid and can move on to the next project.
\end{enumerate}

In the first scenario, the important element to note is that no matter which choice is made, it does not affect the timing of future choices. Denise's next decision is always made the next evening. The time frame is independent of the choice, so we say it is {\it fixed}.

In the second scenario, the time frame depends on the choice made. The timing of Fiona's next choice is determined by her current decision, \eg choosing a shorter project frees her sooner for the next opportunity. We call this the {\it adaptive} time frame.

In our model, we must compute the growth rate of wealth using the time period over which the growth rate is effective. This is the time period between successive decisions. With a fixed time frame it is the period from decision to later payment, \ie $t_b-t_0=\hor+\del$. In Denise's case, this is one day. With an adaptive time frame, it is the period from decision to chosen payment, \ie $t_a-t_0=\hor$ for option $a$ and $t_b-t_0=\hor+\del$ for option $b$. This specification is appropriate to Fiona's situation.

\subhead{Dynamics}
As described in~\sref{model}, the wealth dynamics can also take different forms. We address two common cases: additive and multiplicative wealth dynamics. We note that under the multiplicative dynamics it is assumed that the payment itself is reinvested at a background exponential growth rate, $r$. For additive dynamics there is essentially no reinvestment of the payment. Income in this dynamic is independent of wealth.

We discuss the four specifications, as illustrated in \fref{tree}. In each case we: compute the growth rates $g_a$ and $g_b$ associated with each option; compare them to determine the conditions under which each option is preferred; determine whether preference reversal is predicted; and, finally, elicit the form of temporal discounting equivalent to our decision model.

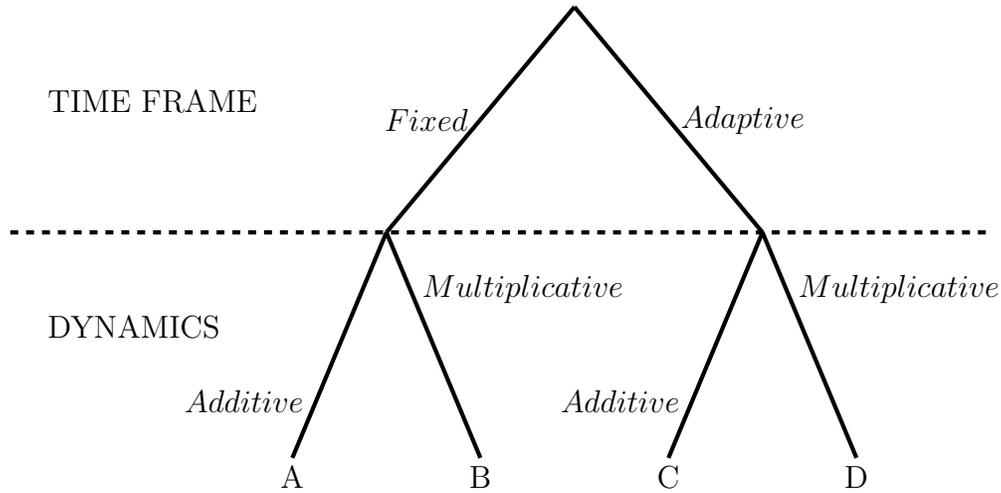
\begin{figure}[!htb]
\centering
\begin{tikzpicture}
\draw[-,ultra thick] (0,6)--(-2.5,3) node[left,midway]{$Fixed$};
\draw[-,ultra thick] (0,6)--(2.5,3) node[right,midway]{$Adaptive$};
\node[text width=3cm] at (-5.5,4.75) {TIME FRAME};
\draw[-,ultra thick,dashed] (-7.5,3)--(5.5,3);
\draw[-,ultra thick] (-2.5,3)--(-3.75,0) node[left,near end]{$Additive$};
\draw[-,ultra thick] (-2.5,3)--(-1.25,0) node[right,near start]{$Multiplicative$};
\draw[-,ultra thick] (2.5,3)--(1.25,0) node[left,near end]{$Additive$};
\draw[-,ultra thick] (2.5,3)--(3.75,0) node[right,near start]{$Multiplicative$};
\node[text width=3cm] at (-5.5,1.75) {DYNAMICS};
\node at (-3.75,-0.25) {A};
\node at (-1.25,-0.25) {B};
\node at (1.25,-0.25) {C};
\node at (3.75,-0.25) {D};
\end{tikzpicture}
\caption{The four model specifications, determined by specifying a time frame and wealth dynamics. The labels A, B, C, and D, are used for the different cases.}
\flabel{tree}
\end{figure}

\subsection{Case A -- Fixed time frame with additive dynamics}\label{sec:case_A}

Specification: the period for computing the growth rate is that between the decision and the later payment, $t_b-t_0=\hor+\del$; and wealth dynamics are additive. Here, irrespective of the initial wealth and in addition to the chosen payment, wealth grows at a background additive growth rate, $k$. In other words, the decision maker always receives an amount $k\left(t_b-t_0\right)$ over the period, on top of what else she chooses.

In the economic context, we might interpret this sinecure-like income as coming from a job or welfare scheme which does not interact with the payments under consideration. In biological contexts, we might interpret negative $k$ as a metabolic rate of energy expenditure. Provided it takes the same value under the two choices, $k$ appears in neither preference criterion nor discount function. We include it only for completeness.

We begin by writing down the final wealth under each of the two options, evaluated at $t_b$:
\bea
x_a\left(t_b\right) &=& x\left(t_0\right) + k\left(t_b-t_0\right) + \Dx_a\,;\\
x_b\left(t_b\right) &=& x\left(t_0\right) + k\left(t_b-t_0\right) + \Dx_b\,.
\eea

The growth rates are:
\bea
g_a &=& \frac{x_a\left(t_b\right) - x\left(t_0\right)}{t_b-t_0} = \frac{\Dx_a}{\hor+\del} + k\,;\\
g_b &=& \frac{x_b\left(t_b\right) - x\left(t_0\right)}{t_b-t_0} = \frac{\Dx_b}{\hor+\del} + k\,.
\eea

Since $\Dx_b > \Dx_a$, option $b$ is always preferred to option $a$. This is a trivial case: under additive wealth dynamics and comparing growth rates over the same time period, only payment size matters. There is no discounting and the discount function $\delta$ is undefined, because the indifference condition is never satisfied.

\subsection{Case B -- Fixed time frame with multiplicative dynamics}\label{sec:case_B}

Specification: the period for computing the growth rate is that between the decision and the later payment, $t_b-t_0=\hor+\del$; and wealth dynamics are multiplicative. This specification corresponds to the classical temporal discounting, where wealth compounds continuously at the background multiplicative growth rate, $r$, and payments are reinvested at this rate.

We note that in this case the earlier payment, $\Dx_a$, if chosen, is treated as growing exponentially from its receipt at $t_a$ to $t_b$. Therefore, the wealths evolve from $t_0$ to $t_b$ as follows:
\bea
x_a\left(t_b\right) &=& x\left(t_0\right) e^{r\left(t_b-t_0\right)} + \Dx_a e^{r\left(t_b-t_a\right)}\,;\\
x_b\left(t_b\right) &=& x\left(t_0\right) e^{r\left(t_b-t_0\right)} + \Dx_b\,.
\eea

The corresponding growth rates are:\footnote{Note that the payments, $\Dx_a$ and $\Dx_b$, appear in the growth rates, $g_a$ and $g_b$, only as their ratios to the initial wealth, $x\left(t_0\right)$. An equivalent choice problem to a decision maker with different wealth has payments scaled up or down, such that these ratios are unchanged. The same comment applies to case D.}
\bea
\elabel{grates_caseB_a}
g_a &=& \frac{\ln x_a\left(t_b\right) - \ln x\left(t_0\right)}{t_b-t_0} = \frac{1}{\hor +\del}\ln{\left(1 + \frac{\Dx_a e^{r\del}}{x\left(t_0\right)e^{r\left(\hor+\del\right)}}\right)} + r\,;\\
\elabel{grates_caseB_b}
g_b &=& \frac{\ln x_b\left(t_b\right) - \ln x\left(t_0\right)}{t_b-t_0} = \frac{1}{\hor +\del}\ln{\left(1 + \frac{\Dx_b}{x\left(t_0\right)e^{r\left(\hor+\del\right)}}\right)} + r\,.
\eea

The criterion $g_a > g_b$ is simple. Only the numerator in the second term of the logarithm is different, so only this must be compared. Thus, $g_a > g_b$ if
\be
\Dx_a e^{r\del} > \Dx_b\,.
\ee

We see that the decision criterion depends on a single time period, the delay $\del$, and on the background growth rate of wealth, $r$. The discount function is obtained by setting the growth rates to be equal, which happens when $\Dx_a e^{r\del} = \Dx_b$. This yields
\be
\delta\left(\del;r\right) = \left.\frac{\Dx_a}{\Dx_b}\right|_{g_a=g_b} = e^{-r\del}\,,
\ee
which is the classical exponential discounting result \citep{Samuelson1937}.\footnote{Indeed, this result corresponds to the historical use of the term ``rate of discount'' to describe a riskless interest rate in the money market, \eg \citet{Jevons1863}.} The interpretation is straightforward: if it is possible to reinvest the earlier payment such that it will exceed the later payment amount at the later payment time, then option $a$ is preferred to option $b$.

\subsection{Case C -- Adaptive time frame with additive dynamics}\label{sec:case_C}

Specification: the period for computing the growth rate is that between the decision and the chosen payment, either $t_a-t_0=\hor$ or $t_b-t_0=\hor+\del$; and wealth dynamics are additive. Like in case A, regardless of the initial wealth and in addition to the chosen payment, wealth grows at the background additive rate, $k$. Unlike case A, options $a$ and $b$ are evaluated at $t_a$ and $t_b$, respectively:
\bea
x_a\left(t_a\right) &=& x\left(t_0\right) + k\left(t_a-t_0\right) + \Dx_a\,;\\
x_b\left(t_b\right) &=& x\left(t_0\right) + k\left(t_b-t_0\right) + \Dx_b\,.
\eea

The growth rates are:
\bea
g_a &=& \frac{x_a\left(t_a\right) - x\left(t_0\right)}{t_a-t_0} = \frac{\Dx_a}{\hor} + k\,;\\
g_b &=& \frac{x_b\left(t_b\right) - x\left(t_0\right)}{t_b-t_0} = \frac{\Dx_b}{\hor +\del} + k\,.
\elabel{grates_caseC}
\eea

It follows that the criterion $g_a > g_b$ is
\be
\frac{\Dx_a}{\hor} > \frac{\Dx_b}{\hor +\del}\,.
\elabel{criterion_caseC}
\ee
So, in this specification, the decision maker cares about the linear payment rate under each option. 

Preference reversals are observed changes in decisions as time passes, \ie as the horizon gets shorter. We can test whether they are predicted in our model by varying $\hor$ while holding other variables constant. In the present specification, indifference occurs at the horizon for which \eref{criterion_caseC} becomes an equality, \ie at $\hor=\hor^\text{PR}$ where
\be
\hor^\text{PR} \equiv \frac{\del \Dx_a}{\Dx_b - \Dx_a}\,.
\elabel{HPR}
\ee
Since $t_0=t_a-\hor$, this can be expressed as a critical decision time at which the decision maker is indifferent:
\be
t_0^\text{PR} \equiv t_a - \hor^\text{PR} = \frac{\Dx_b t_a - \Dx_a t_b}{\Dx_b - \Dx_a}\,.
\elabel{t0PR}
\ee

For $\hor<\hor^\text{PR}$ ($t_0>t_0^\text{PR}$), the payment rate under option $a$ exceeds that under option $b$ and the earlier payment is preferred. The converse is true for $\hor>\hor^\text{PR}$ ($t_0<t_0^\text{PR}$). \fref{caseC} illustrates how the dependence of payment rate on horizon leads to preference reversal under additive dynamics with an adaptive time frame, \cf \citep[Fig.~2]{Kacelnik1997}.

\begin{figure}[!htb]
\centering
\begin{tikzpicture}
\node[inner sep=0pt] at (0,0)
{\includegraphics[width=1.0\textwidth]{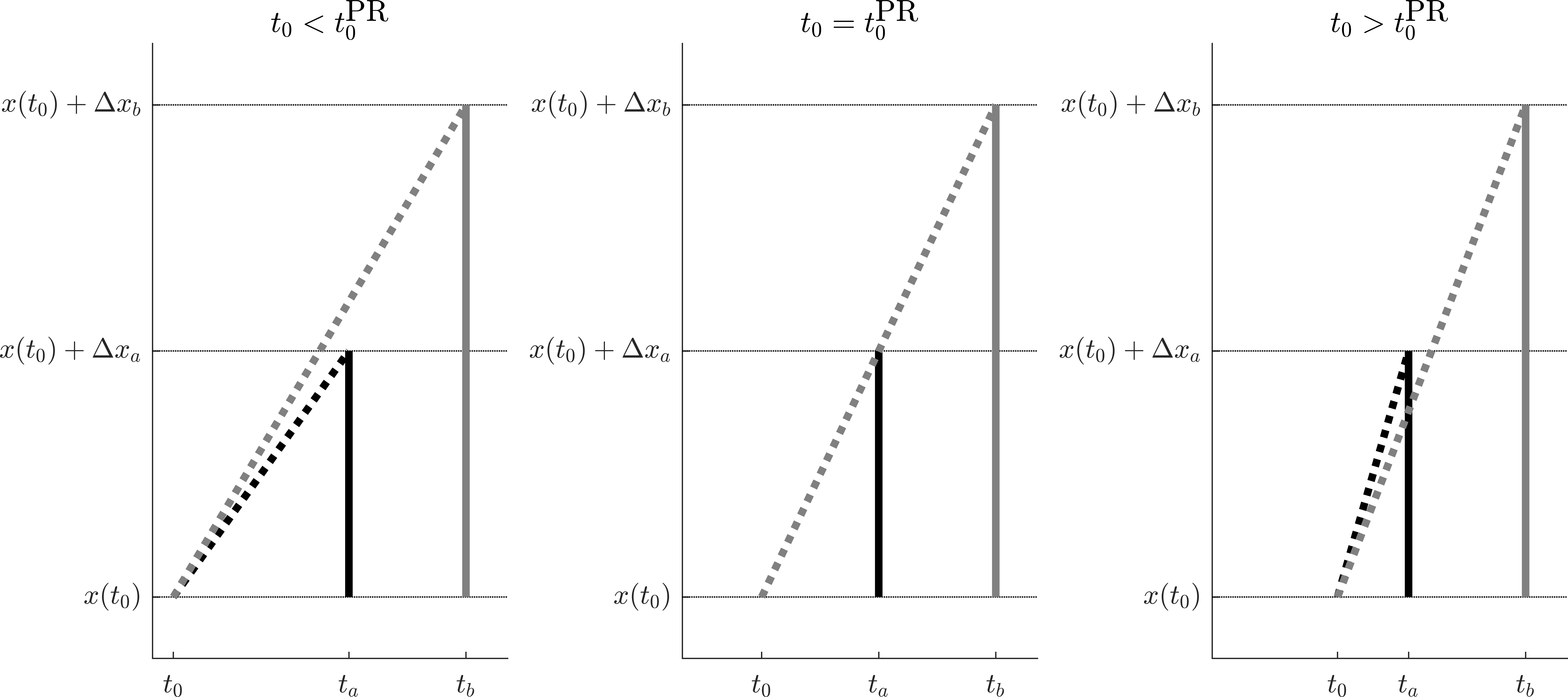}};
\draw [decorate,decoration={brace,amplitude=4pt,mirror},xshift=0pt,yshift=-5pt](-5.9,-2.9) -- (-4.3,-2.9) node [black,midway,yshift=-10pt]{\scriptsize $\hor$};
\draw [decorate,decoration={brace,amplitude=4pt,mirror},xshift=0pt,yshift=-5pt](-4.16,-2.9) -- (-3.16,-2.9) node [black,midway,yshift=-10pt]{\scriptsize $\del$};
\draw [decorate,decoration={brace,amplitude=4pt,mirror},xshift=0pt,yshift=-5pt](-0.15,-2.9) -- (0.85,-2.9) node [black,midway,yshift=-10pt]{\scriptsize $\hor$};
\draw [decorate,decoration={brace,amplitude=4pt,mirror},xshift=0pt,yshift=-5pt](0.99,-2.9) -- (1.99,-2.9) node [black,midway,yshift=-10pt]{\scriptsize $\del$};
\draw [decorate,decoration={brace,amplitude=4pt,mirror},xshift=0pt,yshift=-5pt](5.45,-2.9) -- (6.00,-2.9) node [black,midway,yshift=-10pt]{\scriptsize $\hor$};
\draw [decorate,decoration={brace,amplitude=4pt,mirror},xshift=0pt,yshift=-5pt](6.14,-2.9) -- (7.14,-2.9) node [black,midway,yshift=-10pt]{\scriptsize $\del$};
\end{tikzpicture}
\caption{Preference reversal in case C. From left to right panel, $t_0$ increases and $\hor$ decreases -- \ie both payments get closer -- while all other parameters are held constant. Initially, option $b$ is preferred, having the higher payment rate (slope of dashed line). At the critical time, $t_0=t_0^\text{PR}$, given by \eref{t0PR}, both options imply the same payment rate. At later times, option $a$ has the higher payment rate and is preferred.}
\flabel{caseC}
\end{figure}
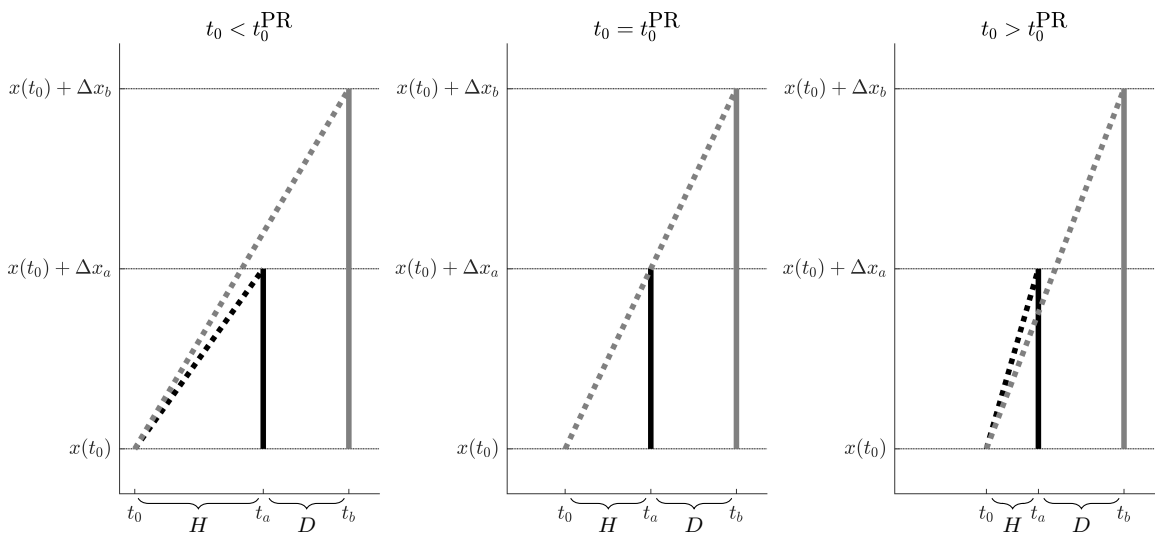

Finally, we compute the discount function under this specification. When $g_a=g_b$, we have
\be
\delta\left(\del;\hor\right) = \left.\frac{\Dx_a}{\Dx_b}\right|_{g_a=g_b} = \frac{\hor}{\hor +\del} = \frac{1}{1+\del/\hor}\,.
\ee
Thus we recover the widely-used descriptive model of discounting in which the discount function, $\delta$, is a hyperbola of the delay, $\del$. We note that $\delta$ also depends on the horizon, $\hor$. Indeed, the degree of discounting parameter -- usually treated as a psychological parameter -- appears in our model as $1/\hor$, the reciprocal of the horizon \citep{MyersonGreen1995}. As the horizon gets shorter, $1/\hor$ becomes larger, $\delta$ gets smaller, and the later payment becomes less favorable. No knowledge of the decision maker's psychology is required in this setup, only the postulate that she prefers her wealth to grow faster rather than slower.



\subsection{Case D -- Adaptive time frame with multiplicative dynamics}\label{sec:case_D}

Specification: the period for computing the growth rate is that between the decision and the chosen payment, either $t_a-t_0=\hor$ or $t_b-t_0=\hor+\del$; and wealth dynamics are multiplicative.

We follow the same steps as in the previous cases. Wealth evolves to:
\bea
x_a\left(t_a\right) &=& x\left(t_0\right) e^{r\left(t_a-t_0\right)} + \Dx_a\,;\\
x_b\left(t_b\right) &=& x\left(t_0\right) e^{r\left(t_b-t_0\right)} + \Dx_b\,.
\eea
The corresponding growth rates are:
\bea
g_a &=& \frac{\ln x_a\left(t_a\right) - \ln x\left(t_0\right)}{t_a-t_0} = \frac{1}{\hor}\ln{\left(1 + \frac{\Dx_a}{x\left(t_0\right)e^{r\hor}}\right)} + r \elabel{ga_D}\,;\\
g_b &=&\frac{\ln x_b\left(t_b\right) - \ln x\left(t_0\right)}{t_b-t_0} = \frac{1}{\hor + \del}\ln{\left(1 + \frac{\Dx_b}{x\left(t_0\right)e^{r\left(\hor + \del\right)}}\right)} + r\,.
\elabel{gb_D}
\eea

\subhead{Preference reversal}
When the later payment is sufficiently large, $\Dx_b > \Dx_a e^{r\del}$, preference reversal is predicted,\footnote{This can be shown by comparing the $\hor\to0$ and $\hor\to\infty$ limits of $g_a$ and $g_b$ in \eref{ga_D} and \eref{gb_D}.} and a threshold horizon, $\hor^\text{PR}$, exists. For shorter horizons than $\hor^\text{PR}$, the earlier payment is preferred ($g_a>g_b$) and \textit{vice versa}. The discount function and threshold horizon are not expressible in closed form for general parameter values. They become tractable in the limit of small payments, which we present below. If the later payment is too small, $\Dx_b < \Dx_a e^{r\del}$, the earlier payment is always preferred in this specification.

\subhead{Wealth effect}
Our model predicts another type of preference reversal here, elicited by varying the initial wealth, $x\left(t_0\right)$, rather than the horizon. As $x\left(t_0\right)\to0$, the earlier payment is preferred regardless of the size of the later payment. If the later payment is large enough, specifically if
\be
\Dx_b > \Dx_a e^{r\del}\left(\frac{\hor+\del}{\hor}\right)\,,
\ee
then it becomes preferable to the earlier payment as $x\left(t_0\right)\to\infty$. Thus, the decision maker switches from preferring earlier to later payments as her wealth increases. We call this the \textit{wealth effect}. It is illustrated pictorially in \fref{reversal_B}, which shows the variation of growth rates, $g_a$ and $g_b$, for each option as initial wealth, $x\left(t_0\right)$, increases from left to right. Other parameters are held constant.

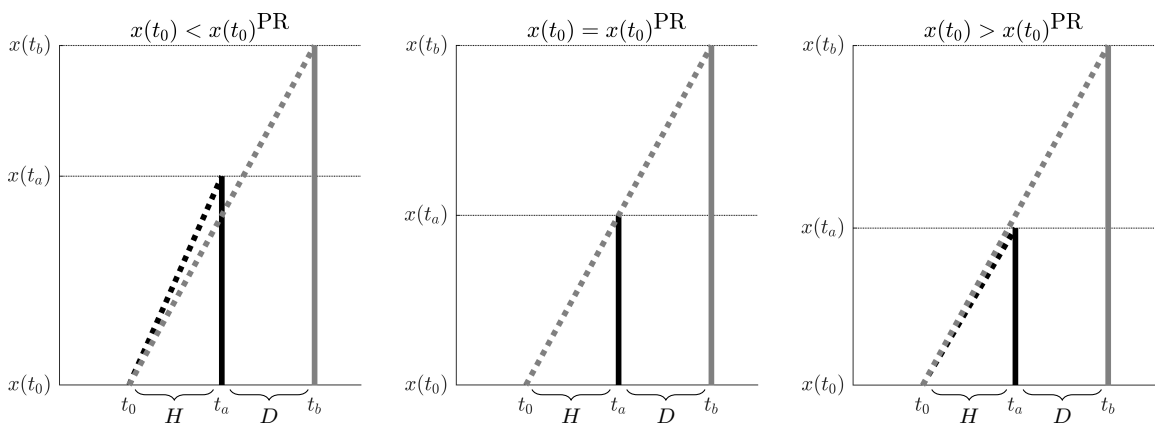
\begin{figure}[!htb]
\centering
\begin{tikzpicture}
\node[inner sep=0pt] at (0,0)
{\includegraphics[width=1.0\textwidth]{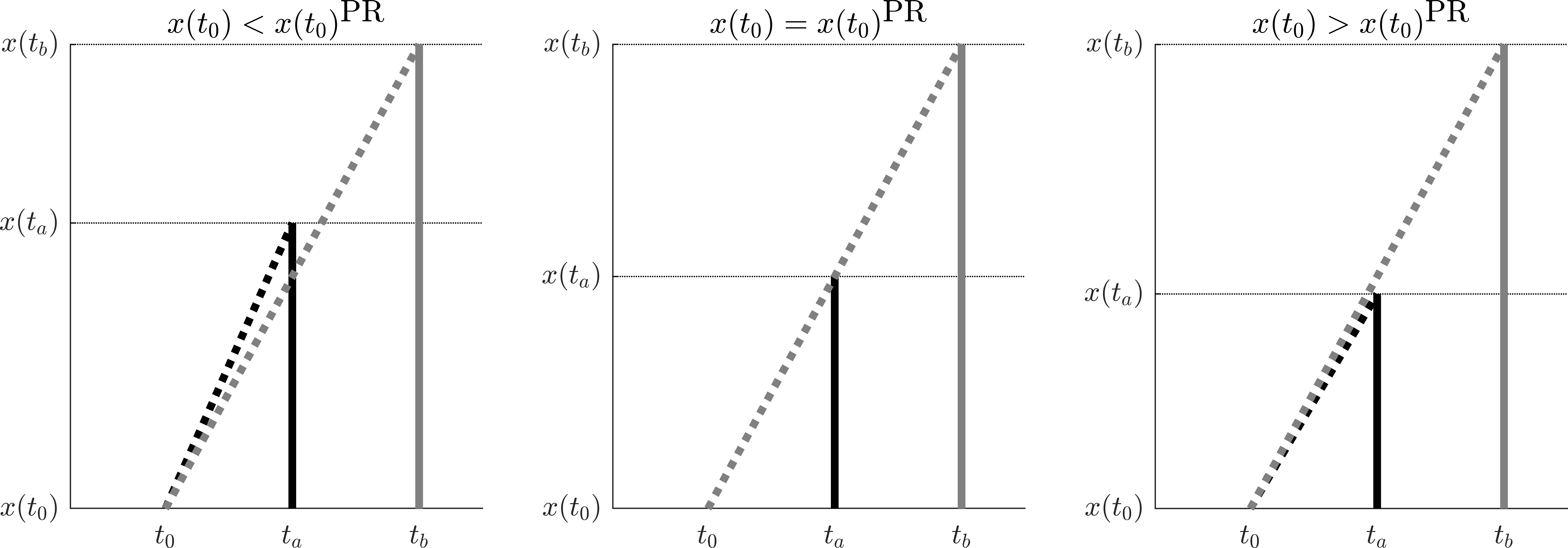}};
\draw [decorate,decoration={brace,amplitude=4pt,mirror},xshift=0pt,yshift=-5pt](-5.94,-2.17) -- (-4.9,-2.17) node [black,midway,yshift=-10pt]{\scriptsize $\hor$};
\draw [decorate,decoration={brace,amplitude=4pt,mirror},xshift=0pt,yshift=-5pt](-4.68,-2.17) -- (-3.64,-2.17) node [black,midway,yshift=-10pt]{\scriptsize $\del$};
\draw [decorate,decoration={brace,amplitude=4pt,mirror},xshift=0pt,yshift=-5pt](-0.665,-2.17) -- (0.375,-2.17) node [black,midway,yshift=-10pt]{\scriptsize $\hor$};
\draw [decorate,decoration={brace,amplitude=4pt,mirror},xshift=0pt,yshift=-5pt](0.595,-2.17) -- (1.635,-2.17) node [black,midway,yshift=-10pt]{\scriptsize $\del$};
\draw [decorate,decoration={brace,amplitude=4pt,mirror},xshift=0pt,yshift=-5pt](4.65,-2.17) -- (5.645,-2.17) node [black,midway,yshift=-10pt]{\scriptsize $\hor$};
\draw [decorate,decoration={brace,amplitude=4pt,mirror},xshift=0pt,yshift=-5pt](5.865,-2.17) -- (6.905,-2.17) node [black,midway,yshift=-10pt]{\scriptsize $\del$};
\end{tikzpicture}
\caption{Wealth effect in case D, with logarithmic vertical scales. Initial wealth $x\left(t_0\right)$ increases from left panel to right panel (\$500, \$2277, \$5500) with all other parameters held fixed ($t_0=$ today, $t_a=1$~year from today, $t_b=2$~years from today, $\Dx_a=\$1000$, $\Dx_b=\$2500$, $r=0.03$~per year). At small wealth, option $a$ is preferred, having the higher growth rate according to \eref{ga_D} and \eref{gb_D}. At a larger wealth, $x\left(t_0\right)^{PR}\approx \$2277$, both options imply equal growth, with reversal occurring as wealth increases further. }
\flabel{reversal_B}
\end{figure}

\Fref{reversal_B2} shows the difference in growth rates, $g_a-g_b$, as a function of initial wealth, $x\left(t_0\right)$, for the same parameters as in \fref{reversal_B}. The earlier payment is preferred when this difference is positive, which happens for wealth below some threshold. For larger wealth, the growth rate difference is negative and the later payment is chosen.

\begin{figure}[!htb]
\centering
\includegraphics[width=0.7\textwidth]{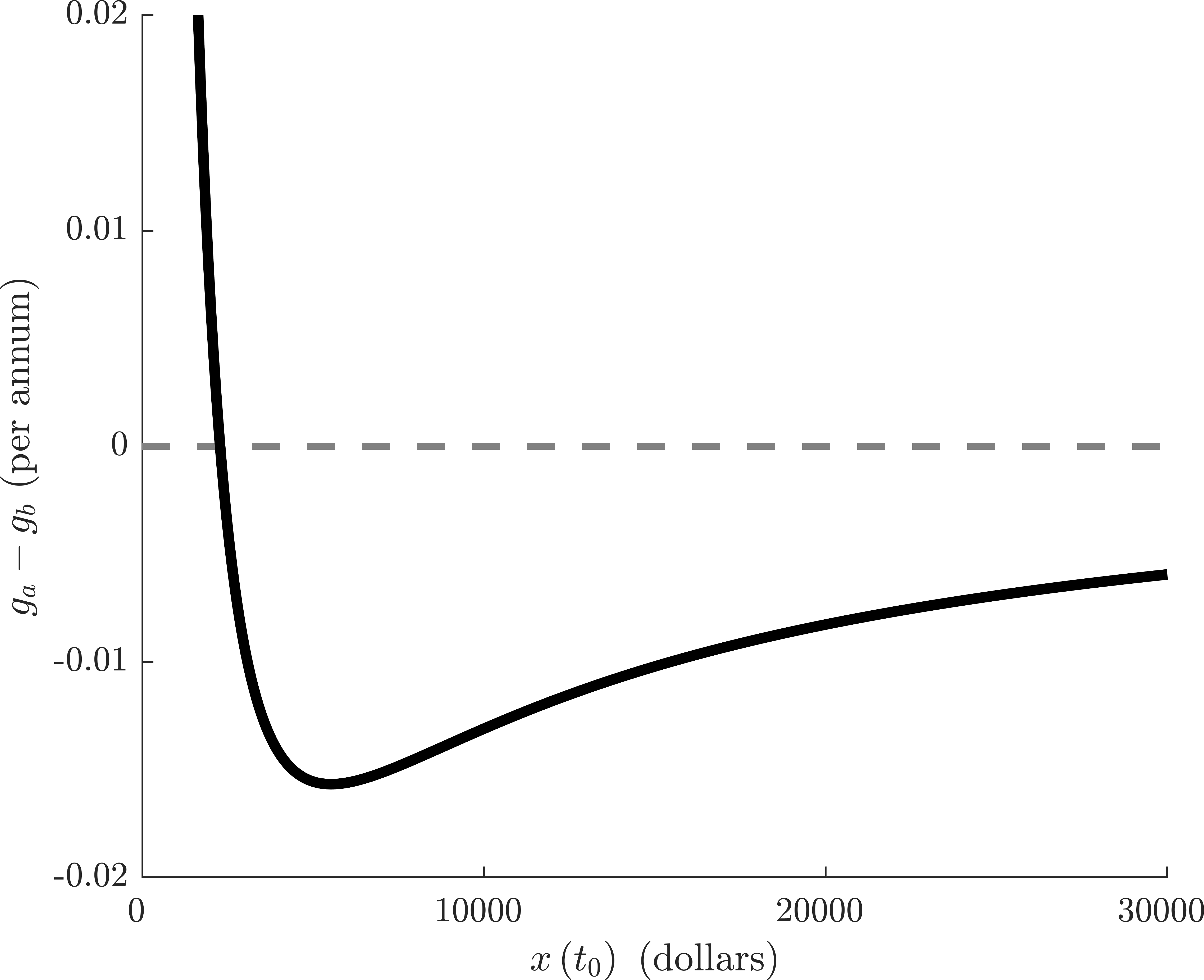}
\caption{The difference in growth rates, $g_a-g_b$, as a function of initial wealth, $x\left(t_0\right)$, in case D. For small initial wealth the earlier, smaller payment is preferred, whereas for large initial wealth the later, larger payment is preferred. Parameters as used in \fref{reversal_B}.}
\flabel{reversal_B2}
\end{figure}

We interpret this as follows. Assuming multiplicative dynamics and an adaptive time frame, it is growth-optimal for people of lower wealth to choose an earlier, smaller payment; and growth-optimal for wealthier individuals to hold out for the later, larger payment. This is consistent with the findings of \citet{EpperETAL2018}, that ``individuals with relatively low time discounting are consistently positioned higher in the wealth distribution.'' It is likely consistent with \citep{GreenETAL1996}, in which people with higher incomes were observed to discount less steeply. 

We exemplify the wealth effect by presenting a calculation using the same parameters as in \fref{reversal_B}. Suppose a decision maker faces a choice between receiving $\$1000$ after one year (option $a$) or $\$2500$ after two years (option $b$), and that she has access to a riskless interest rate of 0.03 per year. If she has $\$500$ initially, she evaluates the growth rate corresponding to option $a$ as
\be
g_a = \frac{1}{1}\ln{\left(1 + \frac{1000}{500e^{0.03\times 1}}\right)} + 0.03 \approx 1.1\text{~per year}\,,
\ee
and to option $b$ as
\be
g_b = \frac{1}{2}\ln{\left(1 + \frac{2500}{500e^{0.03\times 2}}\right)} + 0.03 \approx 0.9\text{~per year}\,.
\ee
Thus, the decision maker would prefer the earlier, smaller payment, as $1.1 > 0.9$. If we assume that the decision maker had initially $\$5500$, \ie, 11 times more than in the previous setting, a similar calculation yields $g_a\approx0.19\text{~per year}$ and $g_b\approx0.21\text{~per year}$, so the later, larger payment is preferred.

\subhead{Discounting in the small payment limit}
In many applications of discounting, it is plausible to assume that the payments are small relative to wealth:
\bea
\Dx_a &\ll& x\left(t_0\right)e^{r\hor}\,;\\
\Dx_b &\ll& x\left(t_0\right)e^{r\left(\hor + \del\right)}\,.
\eea
We can express the threshold horizon and the discount function in closed form in this limit. Setting $g_a=g_b$ and using the first-order approximation $\ln(1+\epsilon)\approx\epsilon$ for $\epsilon\ll1$, we get
\be
\hor^\text{PR} \equiv \frac{\del \Dx_a e^{r \del}}{\Dx_b - \Dx_a e^{r \del}}\,,
\elabel{HPRD}
\ee
and
\be
\delta\left(\del;\hor;r\right) = \left.\frac{\Dx_a}{\Dx_b}\right|_{g_a=g_b} \approx \frac{\hor e^{r\hor}}{\left(\hor + \del\right)e^{r\left(\hor + \del\right)}} = \frac{e^{-r\del}}{1+\del/\hor}\,,
\ee
which is a product of hyperbolic and exponential discount functions. This hybrid case has interesting behavior in the long- and short-delay limits. As shown in \fref{shortpaymentasymp}, discounting is close to hyperbolic for short delays and to exponential for long delays. Thus, for the same dynamic and the same time frame, and assuming small payments relative to initial wealth, our choice criterion predicts both approximately hyperbolic and approximately exponential discounting, depending on the delay.

\begin{figure}[!htb]
\centering
\includegraphics[width=0.7\textwidth]{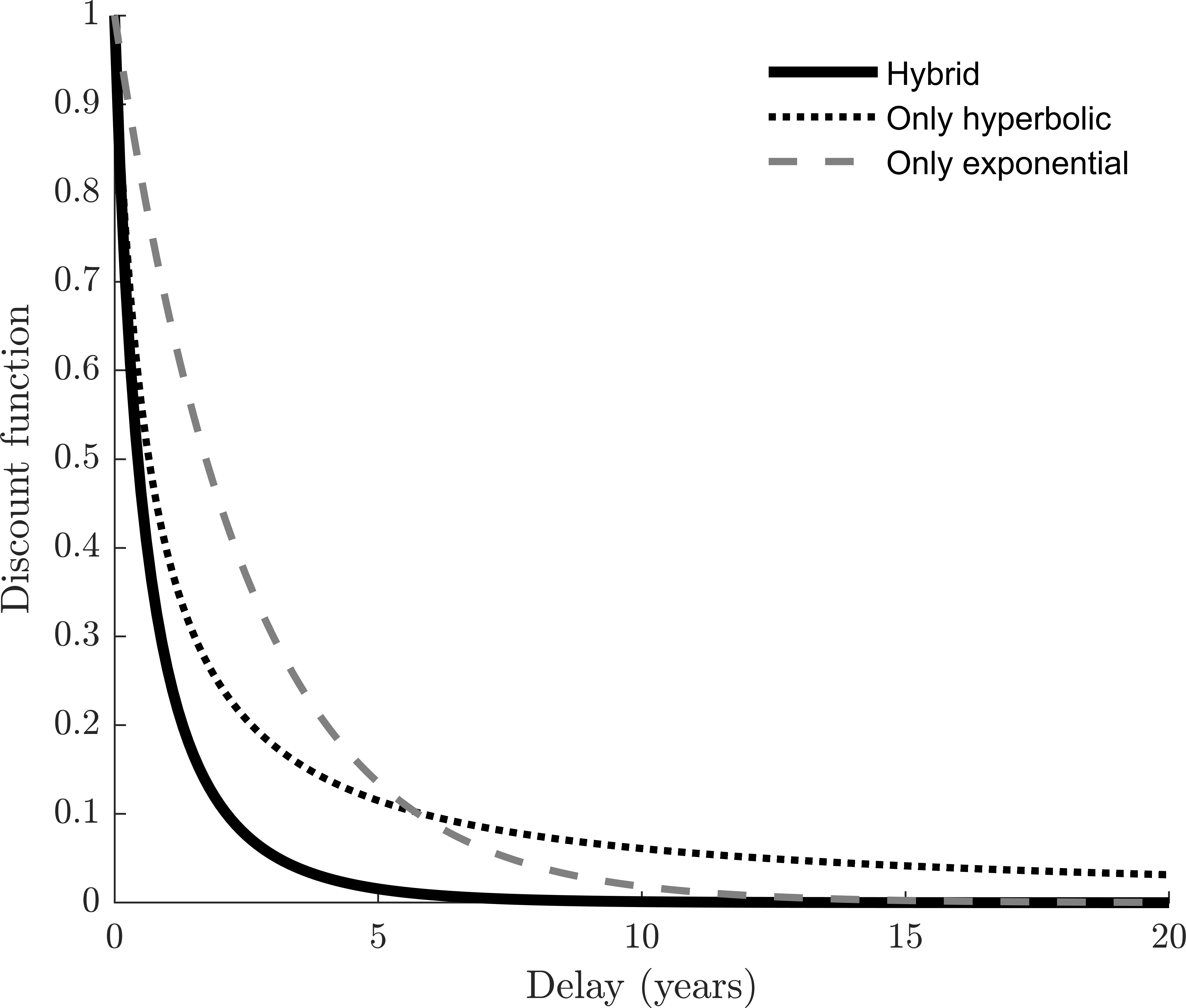}
\caption{The discount function in the small payment limit in case D. The solid black curve is the hybrid $\delta = e^{-r\del}/(1+\del/\hor)$, for $r=0.4\text{ per year}$ and $\hor=0.65\text{ years}$. This is close to the hyperbolic discount function $1/(1+\del/\hor)$ (black dotted) for short delays and to the exponential discount function $e^{-r\del}$ (grey dashed) for long delays.}
\flabel{shortpaymentasymp}
\end{figure}


\section{Discussion}\label{sec:discussion}

This paper explores temporal discounting under the postulate that decision makers maximize the growth rate of their wealth. We consider a basic temporal choice problem between two known, certain, and different payments at known, certain, and different future times. To compute growth rates, the problem must be further specified. We add information about wealth dynamics, treating additive and multiplicative cases, and the time frame of the decision, meaning the period over which growth is evaluated.

Preference reversal is an observed behavior in which decision makers switch from preferring later to earlier payments as time passes. It is incompatible with the classical normative model of exponential discounting. Our model generates four different forms of discounting, depending on the decision maker's circumstances -- no discounting, exponential, hyperbolic, and a hybrid of exponential and hyperbolic. The hyperbolic and hybrid forms predict preference reversal without, as is commonly needed, assumptions of behavioral bias or payment risk.

The hybrid case -- corresponding to multiplicative dynamics and an adaptive time frame -- suggests another type of preference reversal, called the wealth effect. Here a decision maker switches from an earlier to a later payment as her wealth increases. In other words, richer people discount less steeply than poorer people, in line with empirical findings \citep{GreenETAL1996,EpperETAL2018}.

Our main contribution is the prediction of non-exponential discounting and preference reversal in a model that does not violate standard axioms of choice \citep{vonNeumannMorgenstern1944}. Changes in the discount function arise only from changes in wealth dynamics and time frame. This marks a shift from psychological to circumstantial explanations of discounting. Our model assumes no dynamic inconsistency, in that the decision maker prefers at all times the option with the highest growth rate. If corroborated empirically, it would be both a normative and a descriptive model. Experimental tests are feasible because the model works directly with money payments rather than utility-of-consumption flows \citep{CohenETAL2019}.

The temporal choice problem we study is riskless. A planned extension of this work is to explore the consequences for our model of payment uncertainty. Uncertainty decreases the long-time growth rate associated with a payment \citep{PetersGell-Mann2016}. This would make a risky payment less desirable in our model, without reference to the risk preferences of the decision maker.

The dynamics discussed here do not cover the entire range of wealth dynamics. Although multiplicative and additive wealth dynamics are common and intuitive, other wealth dynamics are possible, which would lead to other forms of discounting in our model. Our decision criterion can be adapted to general dynamics using the growth rates described in \citep{PetersAdamou2018a}. 



We end with a caveat. The temporal choice problem involves two {\it future} payments. In the small horizon limit, the growth rate corresponding to the earlier payment in the adaptive time frame diverges. This indicates a loss of model realism. We link this to the breakdown of an implicit assumption: that the growth rates we compute are sustained over sufficiently long periods to be meaningful to the decision maker. They are, in effect, the growth rates of wealth achieved under repetition of the choice. We share the view of \citet[p.~60]{Kacelnik1997} that ``the discounting process used for the one-off events seems to obey a law that evolved as an adaptation to cope with repetitive events.'' As the horizon shrinks, this imagined repetition occurs at a frequency so high that the choice problem no longer resembles a real situation.

\bibliography{./bibliography}

\begin{thebibliography}{19}
\newcommand{\enquote}[1]{``#1''}
\expandafter\ifx\csname natexlab\endcsname\relax\def\natexlab#1{#1}\fi

\bibitem[\protect\citeauthoryear{Cohen, Marzilli~Ericson, Laibson, and
  White}{Cohen et~al.}{2019}]{CohenETAL2019}
\textsc{Cohen, J.~D., K.~M. Marzilli~Ericson, D.~Laibson, and J.~M. White}
  (2019): \enquote{Measuring Time Preferences,} \emph{Journal of Economic
  Literature}, \textit{Forthcoming}.

\bibitem[\protect\citeauthoryear{Dasgupta and Maskin}{Dasgupta and
  Maskin}{2005}]{dasgupta2005uncertainty}
\textsc{Dasgupta, P. and E.~Maskin} (2005): \enquote{Uncertainty and Hyperbolic
  Discounting,} \emph{American Economic Review}, 95, 1290--1299.

\bibitem[\protect\citeauthoryear{Epper, Fehr, Fehr-Duda, Kreiner, Lassen,
  Leth-Petersen, and Rasmussen}{Epper et~al.}{2018}]{EpperETAL2018}
\textsc{Epper, T., E.~Fehr, H.~Fehr-Duda, C.~T. Kreiner, D.~D. Lassen,
  S.~Leth-Petersen, and G.~N. Rasmussen} (2018): \enquote{Time Discounting and
  Wealth Inequality,} .

\bibitem[\protect\citeauthoryear{Green and Myerson}{Green and
  Myerson}{1996}]{GreenMyerson1996}
\textsc{Green, L. and J.~Myerson} (1996): \enquote{Exponential versus
  Hyperbolic Discounting of Delayed Outcomes: Risk and Waiting Time,}
  \emph{American Zoologist}, 36, 496--505.

\bibitem[\protect\citeauthoryear{Green, Myerson, Lichtman, Rosen, and
  Fry}{Green et~al.}{1996}]{GreenETAL1996}
\textsc{Green, L., J.~Myerson, D.~Lichtman, S.~Rosen, and A.~Fry} (1996):
  \enquote{Temporal Discounting in Choice between Delayed Rewards: The Role of
  Age and Income,} \emph{Psychology and Aging}, 11, 79--84.

\bibitem[\protect\citeauthoryear{Holling}{Holling}{1959}]{Holling1959}
\textsc{Holling, C.~S.} (1959): \enquote{Some Characteristics of Simple Types
  of Predation and Parasitism,} \emph{The Canadian Entomologist}, 91, 385--398.

\bibitem[\protect\citeauthoryear{Jevons}{Jevons}{1863}]{Jevons1863}
\textsc{Jevons, W.~S.} (1863): \emph{A Serious Fall in the Value of Gold
  Ascertained, and Its Social Effects Set Forth}, E. Stanford.

\bibitem[\protect\citeauthoryear{Kacelnik}{Kacelnik}{1997}]{Kacelnik1997}
\textsc{Kacelnik, A.} (1997): \enquote{Normative and Descriptive Models of
  Decision Making: Time Discounting and Risk Sensitivity,} in
  \emph{Characterizing Human Psychological Adaptations}, ed. by G.~R. Bock and
  G.~Cardew, John Wiley \& Sons Ltd., 51--70.

\bibitem[\protect\citeauthoryear{Keren and Roelofsma}{Keren and
  Roelofsma}{1995}]{KerenRoelofsma1995}
\textsc{Keren, G. and P.~Roelofsma} (1995): \enquote{Immediacy and Certainty in
  Intertemporal Choice,} \emph{Organizational Behavior and Human Decision
  Processes}, 63, 287--297.

\bibitem[\protect\citeauthoryear{Laibson}{Laibson}{1997}]{Laibson1997}
\textsc{Laibson, D.} (1997): \enquote{Golden Eggs and Hyperbolic Discounting,}
  \emph{Quarterly Journal of Economics}, 112, 443--478.

\bibitem[\protect\citeauthoryear{Loewenstein and Prelec}{Loewenstein and
  Prelec}{1992}]{LoewensteinPrelec1992}
\textsc{Loewenstein, G. and D.~Prelec} (1992): \enquote{Anomalies in
  Intertemporal Choice: Evidence and an Interpretation,} \emph{Quarterly
  Journal of Economics}, 107, 573--597.

\bibitem[\protect\citeauthoryear{Meder, Rabe, Morville, Madsen, Koudahl, Dolan,
  Siebner, and Hulme}{Meder et~al.}{2019}]{MederETAL2019}
\textsc{Meder, D., F.~Rabe, T.~Morville, K.~H. Madsen, M.~T. Koudahl, R.~J.
  Dolan, H.~R. Siebner, and O.~J. Hulme} (2019): \enquote{Ergodicity-Breaking
  Reveals Time Optimal Economic Behavior in Humans,} \emph{ar{X}iv:1906.04652}.

\bibitem[\protect\citeauthoryear{Myerson and Green}{Myerson and
  Green}{1995}]{MyersonGreen1995}
\textsc{Myerson, J. and L.~Green} (1995): \enquote{Discounting of Delayed
  Rewards: Models of Individual Choice,} \emph{Journal of the Experimental
  Analysis of Behavior}, 64, 263--276.

\bibitem[\protect\citeauthoryear{Peters}{Peters}{2019}]{Peters2019b}
\textsc{Peters, O.} (2019): \enquote{The Ergodicity Problem in Economics,}
  \emph{Nature Physics}, 15, 1216--1221.

\bibitem[\protect\citeauthoryear{Peters and Adamou}{Peters and
  Adamou}{2018}]{PetersAdamou2018a}
\textsc{Peters, O. and A.~Adamou} (2018): \enquote{The Time Interpretation of
  Expected Utility Theory,} \emph{ar{X}iv:1801.03680}.

\bibitem[\protect\citeauthoryear{Peters and Gell-Mann}{Peters and
  Gell-Mann}{2016}]{PetersGell-Mann2016}
\textsc{Peters, O. and M.~Gell-Mann} (2016): \enquote{Evaluating Gambles using
  Dynamics,} \emph{Chaos}, 26, 23103.

\bibitem[\protect\citeauthoryear{Samuelson}{Samuelson}{1937}]{Samuelson1937}
\textsc{Samuelson, P.~A.} (1937): \enquote{A Note on Measurement of Utility,}
  \emph{Review of Economic Studies}, 4, 155--161.

\bibitem[\protect\citeauthoryear{Sozou}{Sozou}{1998}]{Sozou1998}
\textsc{Sozou, P.~D.} (1998): \enquote{On Hyperbolic Discounting and Uncertain
  Hazard Rates,} \emph{Proceedings of the Royal Society B: Biological
  Sciences}, 265, 2015--2020.

\bibitem[\protect\citeauthoryear{von Neumann and Morgenstern}{von Neumann and
  Morgenstern}{1944}]{vonNeumannMorgenstern1944}
\textsc{von Neumann, J. and O.~Morgenstern} (1944): \emph{Theory of Games and
  Economic Behavior}, Princeton University Press.

\end{thebibliography}
\clearpage
\appendix

\section{The Transitivity of Growth-Optimal Preferences}\label{app:appA}

In this appendix we show that growth-optimal preferences satisfy transitivity for all four cases described in the paper. To prove transitivity we assume three payments, $a\equiv\left(t_a,\Dx_a\right)$, $b\equiv\left(t_b,\Dx_b\right)$ and $c\equiv\left(t_c,\Dx_c\right)$, where $t_a < t_b < t_c$. We also assume a decision time $t_0 < t_a$ and an initial wealth $x\left(t_0\right)$. These payments define three RIPPs: a comparison between $a$ to $b$, between $b$ to $c$, and between $a$ to $c$.

In each of the four cases we will show that if $a \prec b$ and $b \prec c$, then $a \prec c$. We will also show that if $a \sim b$ and $b \sim c$, then $a \sim c$.

\subhead{Case A}
In case A (see \sref{case_A}), we show that growth rate maximization is achieved by choosing the larger payment. Therefore, $a \prec b$ iff $\Dx_a < \Dx_b$ and $b \prec c$ iff $\Dx_b < \Dx_c$. It follows that $a \prec c$ because $\Dx_a < \Dx_c$. If $a \sim b$ and $b \sim c$ then $\Dx_a = \Dx_b$ and $\Dx_b = \Dx_c$, so $\Dx_a = \Dx_c$ and therefore $a \sim c$.

\subhead{Case B}
In case B (see \sref{case_B}), we show that growth rate maximization is achieved by comparing the earlier payment to the later payment discounted by an exponential function, so
\bea
a \prec b &\iff& \Dx_a < \Dx_b e^{-r\left(t_b - t_a\right)}\,;\\
b \prec c &\iff& \Dx_b < \Dx_c e^{-r\left(t_c - t_b\right)}\,.
\eea
It follows that $\Dx_b e^{-r\left(t_b - t_a\right)} < \Dx_c e^{-r\left(t_c - t_b\right)} e^{-r\left(t_b - t_a\right)} = \Dx_c e^{-r\left(t_c - t_a\right)}$, so
\be
\Dx_a < \Dx_c e^{-r\left(t_c - t_a\right)} \Longrightarrow a \prec c\,.
\ee
Similarly,
\bea
a \sim b &\iff& \Dx_a = \Dx_b e^{-r\left(t_b - t_a\right)}\,;\\
b \sim c &\iff& \Dx_b = \Dx_c e^{-r\left(t_c - t_b\right)}\,.
\eea
It follows that $\Dx_b e^{-r\left(t_b - t_a\right)} = \Dx_c e^{-r\left(t_c - t_a\right)}$, so
\be
\Dx_a = \Dx_c e^{-r\left(t_c - t_a\right)} \Longrightarrow a \sim c\,.
\ee

\subhead{Case C}
In case C (see \sref{case_C}) only the linear payment rate of each option matters to the decision maker, so
\bea
a \prec b &\iff& \frac{\Dx_a}{t_a - t_0} < \frac{\Dx_b}{t_b - t_0}\,;\\
b \prec c &\iff& \frac{\Dx_b}{t_b - t_0} < \frac{\Dx_c}{t_c - t_0}\,.
\eea
It follows that $\frac{\Dx_a}{t_a - t_0} < \frac{\Dx_c}{t_c - t_0}$, and $a \prec c$. Similarly,
\bea
a = b &\iff& \frac{\Dx_a}{t_a - t_0} = \frac{\Dx_b}{t_b - t_0}\,;\\
b = c &\iff& \frac{\Dx_b}{t_b - t_0} = \frac{\Dx_c}{t_c - t_0}\,,
\eea
so $\frac{\Dx_a}{t_a - t_0} = \frac{\Dx_c}{t_c - t_0}$, and $a \sim c$.

\subhead{Case D}
Like in case C, the time frame in case D (see \sref{case_D}) is adaptive. For this reason the growth rate associated with each payment depends only on the payment time and the decision time. In other words, under both RIPPs comparing $a$ to $b$, and $a$ to $c$, the growth rate associated with payment $a$, $g_a$, is the same. Similarly, $g_b$ is the same in both RIPPs comparing $a$ to $b$, and $b$ to $c$, and $g_c$ is the same in both RIPPs comparing $a$ to $c$, and $b$ to $c$.

It follows that
\bea
a \prec b &\iff& g_a < g_b\,;\\
b \prec c &\iff& g_b < g_c\,,
\eea
so $g_a < g_c$, and $a \prec c$. Similarly,
\bea
a \sim b &\iff& g_a = g_b\,;\\
b \sim c &\iff& g_b = g_c\,,
\eea
so $g_a = g_c$, and $a \sim c$.

\qed

\end{document}